\documentclass[12pt]{iopart}

\usepackage{graphicx} 

\begin{document}

\title[Kagom\'{e} antiferromagnets Cs$_{2}$Cu$_{3}M$F$_{12}$ ($M=$Zr, Hf)] {$\bi{S}\,\mathbf{=1/2}$ Kagom\'{e} antiferromagnets Cs$_{2}$Cu$_{3}$MF$_{12}$ with M$=$Zr and Hf}

\author{Y Yamabe, T Ono, T Suto and H Tanaka}

\address{Department of Physics, Tokyo Institute of Technology\\ 
Oh-okayama, Meguro-ku, Tokyo 152-8551, Japan}
\ead{o-toshio@lee.phys.titech.ac.jp}
\begin{abstract}
Magnetization and specific heat measurements have been carried out on Cs$_2$Cu$_3$ZrF$_{12}$ and Cs$_2$Cu$_3$HfF$_{12}$ single crystals, in which Cu$^{2+}$ ions with spin-$1/2$ form a regular Kagom\'{e} lattice. The antiferromagnetic exchange interaction between neighboring Cu$^{2+}$ spins is $J/k_{\rm B}\simeq 360$ K and 540 K for Cs$_2$Cu$_3$ZrF$_{12}$ and Cs$_2$Cu$_3$HfF$_{12}$, respectively. Structural phase transitions were observed at $T_{\rm t}\simeq 210$ K and 175 K for Cs$_2$Cu$_3$ZrF$_{12}$ and Cs$_2$Cu$_3$HfF$_{12}$, respectively. The specific heat shows a small bend anomaly indicative of magnetic ordering at $T_\mathrm{N}= 23.5$ K and 24.5 K in Cs$_2$Cu$_3$ZrF$_{12}$ and Cs$_2$Cu$_3$HfF$_{12}$, respectively. Weak ferromagnetic behavior was observed below $T_\mathrm{N}$. This weak ferromagnetism should be ascribed to the antisymmetric interaction of the Dzyaloshinsky-Moriya type that are generally allowed in the Kagom\'{e} lattice.
\end{abstract}


\section{Introduction}
Antiferromagnets with strong geometrical frustration are fascinating systems which can display a variety of exotic magnetic phases different from conventional N\'{e}el ordering\,\cite{Collins,Harrison}. Recently, frustrated systems with strong quantum fluctuation have been attracting much attention from a view point of the interplay of quantum fluctuation and spin frustration. $S=1/2$ Kagom\'{e} antiferromagnet (KAF) is a typical example of such systems. There are many theoretical works on the  $S=1/2$ KAF. By virtue of careful analyses and numerical simulation, it has been predicted that the ground state of $S=1/2$ KAF is quantum spin liquid as resonating valence bond state (RVB), and that there are huge number of singlet excitations which fill up the triplet excitation gap\,\cite{Lecheminant,Waldtmann,Mambrini}.  However, this intriguing prediction has not been demonstrated experimentally. The experimental study of the $S=1/2$ KAF is limited because of few model substances.   
Three compounds [Cu$_{3}$(titmb)$_2$(CH$_3$CO$_2$)$_6$]$\cdot$H$_2$O\,\cite{Hiroi}, Cu$_3$V$_2$O$_7$(OH)$_2\cdot$2H$_2$O\,\cite{Rogado} and $\beta$-Cu$_3$V$_2$O$_8$\,\cite{Narumi} are known to have Kagom\'{e} lattice or closely related lattices. However the Kagom\'{e} net is distorted into an orthorhombic form in Cu$_3$V$_2$O$_7$(OH)$_{2}\cdot$2H$_2$O, and buckled like staircase in $\beta$-Cu$_3$V$_2$O$_8$, so that the exchange network is anisotropic. For [Cu$_3$(timb)$_2$(CH$_3$CO$_2$)$_6$] $\cdot$H$_2$O, the nearest neighbor exchange interaction is ferromagnetic.

Cs$_2$Cu$_3$ZrF$_{12}$ is a new candidate of $S=1/2$ KAF\,\cite{Suto}. In this paper, we will report the magnetic properties in Cs$_2$Cu$_3$HfF$_{12}$ in addition to those in Cs$_2$Cu$_3$ZrF$_{12}$, both of which were originally synthesized by M\"{u}ller {\it et al.}\,\cite{Mueller}. Figure 1 shows the room temperature crystal structure of Cs$_2$Cu$_3$MF$_{12}$ (M$=$Zr, Hf) and its projection onto the $c$-plane. These compounds are crystallized in trigonal structures (space group $R\bar{3}m$) \cite{Mueller}. CuF$_6$ octahedra are connected in the $c$-plane sharing corners. Magnetic $\mathrm{Cu^{2+}}$ ions with $S=1/2$ form a regular Kagom\'{e} lattice in the $c$-plane, and all of the nearest neighbor exchange interactions are equivalent. CuF$_6$ octahedra are elongated along the principal axes which are almost parallel to the $c$-axis, so that the hole orbitals $d(x^2-y^2)$ of Cu$^{2+}$ spread in the Kagom\'{e} layer. Since the bond angle of superexchange $\mathrm{Cu^{2+}-F^{-}-Cu^{2+}}$ in the $c$ plane is about $140^{\circ}$, the nearest neighbor superexchange interaction $J$ through F$^-$ ion in the Kagom\'{e} layer should be antiferromagnetic and strong. The interlayer exchange interaction $J'$ should be much smaller than $J$, because magnetic $\mathrm{Cu^{2+}}$ layers are sufficiently separated by nonmagnetic $\mathrm{Cs^{+}}$, $\mathrm{Zr^{4+}}$ ($\mathrm{Hf^{4+}}$) and F$^-$ layers. Thus, the present systems can be described as quasi-two-dimensional (2D) $S=1/2$ KAF.
\begin{figure}[tbp]
	\begin{center}
		\includegraphics[width=15cm, clip]{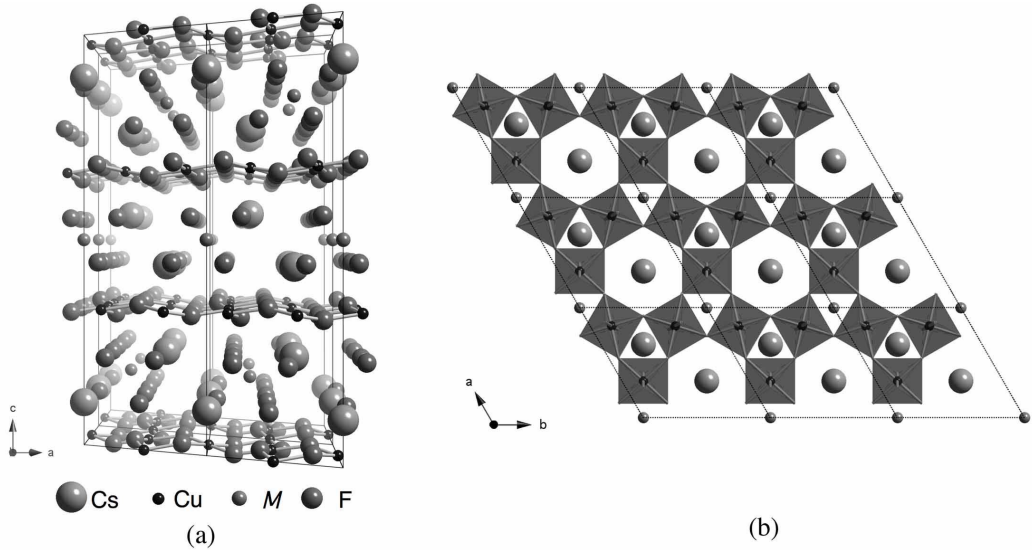}
	\end{center}
	\caption{(a) The crystal structure of Cs$_2$Cu$_{3}M$F$_{12}$ (M$=$Zr, Hf) and (b) its projection onto the $c$-plane. CuF$_6$ octahedra are shaded in (b).}
	\label{layer}
\end{figure}

\section{Experiments}
$\mathrm{Cs_2Cu_3MF_{12}}$ with M$=$ Zr and Hf crystals were synthesized according to the chemical reaction $\mathrm{2CsF + 3CuF_2 + MF_4}$ $\rightarrow$ $\mathrm{Cs_2Cu_3MF_{12}}$. 
$\mathrm{CsF}$, $\mathrm{CuF_2}$ and $\mathrm{ZrF_4}$ were dehydrated by heating in vacuum at ${\sim}$150$^{\circ}$C. The materials were packed into Pt-tube in the ratio of $2:3:1$.
Single crystals of $\mathrm{Cs_2Cu_3MF_{12}}$ were grown by both vertical and horizontal Bridgman methods. The temperature at the center of the furnace was set at 700$^{\circ}$C. Transparent colorless crystals were obtained. The crystals obtained were identified to be $\mathrm{Cs_2Cu_3MF_{12}}$ by X-ray powder and single crystal diffractions. The crystals are cleaved parallel to the $c$-plane. Magnetization was measured down to 1.8 K in magnetic fields up to 7 T, using a SQUID magnetometer (Quantum Design MPMS XL). The specific heat was measured down to 1.8 K, using a Physical Property Measurement System (Quantum Design PPMS) by the relaxation method.

\section{Results and discussion}
Figure 2 shows the inverse magnetic susceptibilities of $\mathrm{Cs_2Cu_3ZrF_{12}}$ and $\mathrm{Cs_2Cu_3HfF_{12}}$ as functions of temperature measured for external field parallel to the $c$-axis. With decreasing temperature, the susceptibilities increase obeying the Curie-Weiss law with large negative Weiss constants, which is indicative of strong antiferromagnetic exchange interaction. The Weiss constants obtained for $H\parallel c$ are listed in Table 1. The susceptibilities exhibit sudden jumps at $T_{\rm t}\simeq 210$ K and 175 K for $\mathrm{Cs_2Cu_3ZrF_{12}}$ and $\mathrm{Cs_2Cu_3HfF_{12}}$, respectively. In both systems, the phase transition at $T_{\rm t}$ has small hysteresis, which is indicative of the first order transition. The susceptibilities obey the Curie-Weiss law also below $T_{\rm t}$. These results indicate that the phase transition at $T_{\rm t}$ is not of magnetic but of structural origin. We tried to analyze the crystal structure below $T_{\rm t}$ by X-ray diffraction but did not succeeded because of multidomain structures. 
\begin{figure}[tbp]
	\begin{center}
		\includegraphics[width=15cm, clip]{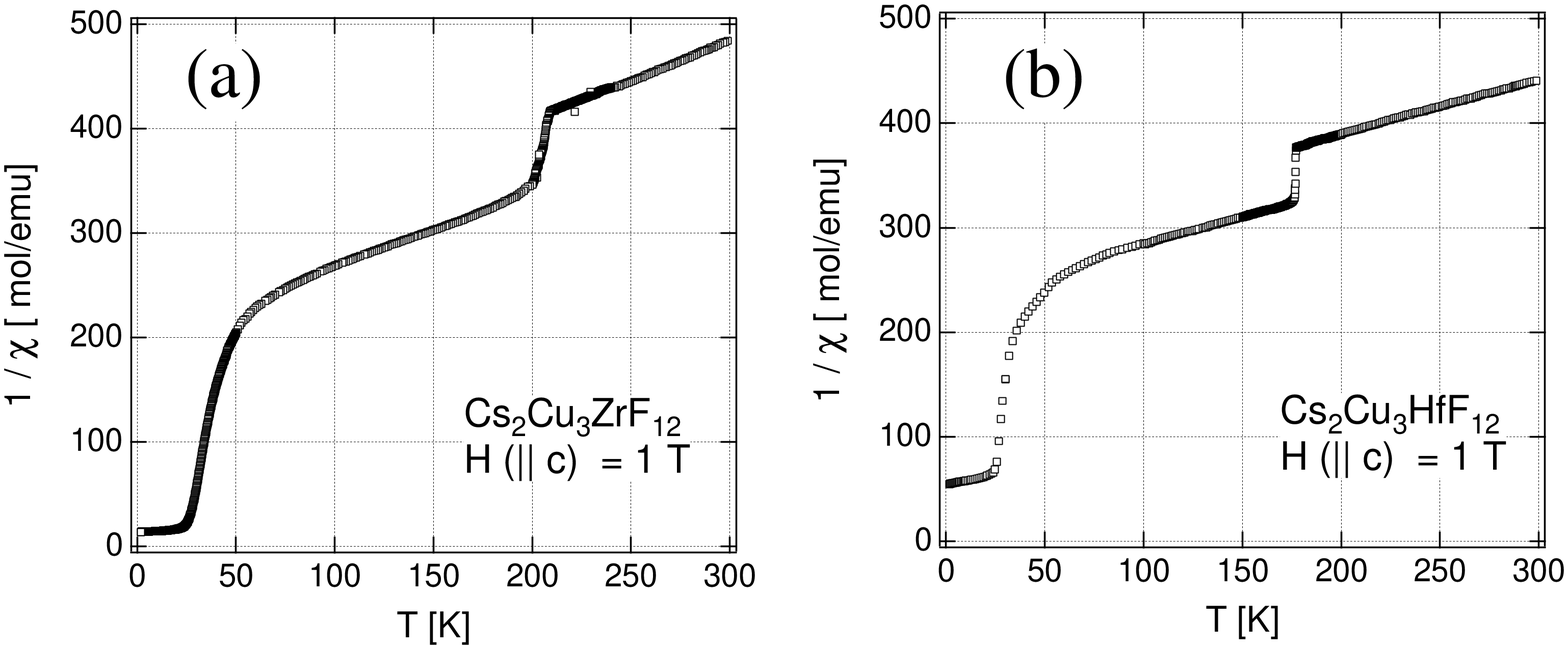}
	\end{center}
	\caption{Temperature dependences of inverse magnetic susceptibilities in (a) $\mathrm{Cs_2Cu_3ZrF_{12}}$ and (b) $\mathrm{Cs_2Cu_3HfF_{12}}$ measured at $H=0.1$ T for $H\parallel c$.}
	\label{chi}
\end{figure}

\Table{\label{Weiss}Structural phase transition temperature $T_{\rm t}$, Weiss constants ${\Theta}$ and spontaneous magnetization $M_{\rm sp}$ at 1.8 K in $\mathrm{Cs_2Cu_3ZrF_{12}}$ and $\mathrm{Cs_2Cu_3HfF_{12}}$, where $T_{\rm t}$ and ${\Theta}$ are in K units and $M_{\rm sp}$ is in $\mu_{\rm B}/{\rm Cu^{2+}}$ unit. ${\Theta}$ is obtained for $H\parallel c$.}
\br
Substance & & $T_{\rm t}$ & & ${\Theta}\ (\,> T_{\rm t})$ &  ${\Theta}\ (\,< T_{\rm t})$ &  $M_{\rm sp}\ (H\parallel c)$ &  $M_{\rm sp}\ (H\perp c)$\\
\mr
$\mathrm{Cs_2Cu_3ZrF_{12}}$ & & 210  & & $-360$ &  $-300$ &  $\simeq 0.015$ &  $\simeq 0.071$ \\ 
$\mathrm{Cs_2Cu_3HfF_{12}}$ & & 175  & & $-540$ &  $-450$ & $\simeq 0.006$ &  $\simeq 0.048$ \\
\br
\endTable

With further decrease of temperature, the magnetic susceptibilities increase rapidly below 40 K for $\mathrm{Cs_2Cu_3ZrF_{12}}$ and below 50 K for $\mathrm{Cs_2Cu_3HfF_{12}}$. The spontaneous magnetizations were observed at $T=1.8$ K, as listed in Table 1. The weak ferromagnetic moment for $H\parallel c$ is much smaller than that for $H\perp c$. The weak ferromagnetic moment is intrinsic to the present system, because its magnitude is strongly dependent on field direction and independent of sample. Therefore, the rapid increase in the magnetic susceptibility at low temperatures is originated from magnetic ordering with the weak ferromagnetic moment.

Within the mean-field theory, the Weiss constant for the Kagom\'{e} lattice is related to the nearest neighbor exchange interaction as ${\Theta}=-J/k_{\rm B}$, where $J$ is defined as ${\cal H}=\sum_{\langle i,j\rangle}J\left(\bi{S}_{i}\cdot \bi{S}_{j}\right)$.
From the Weiss constants listed in Table 1, we obtain $J_{\rm H}/k_{\rm B}\simeq 360$ K and $J_{\rm L}/k_{\rm B}\simeq 300$ K for $\mathrm{Cs_2Cu_3ZrF_{12}}$, and $J_{\rm H}/k_{\rm B}\simeq 540$ K and $J_{\rm L}/k_{\rm B}\simeq 450$ K for $\mathrm{Cs_2Cu_3HfF_{12}}$, where $J_{\rm H}$ and $J_{\rm L}$ are exchange interactions above and below $T_{\rm t}$, respectively. In both systems, the exchange interactions are antiferromagnetic and strong. This is attributed to that the hole orbitals $d(x^2-y^2)$ of Cu$^{2+}$ spread in the Kagom\'{e} layer, and that the bond angle $\mathrm{Cu^{2+}-F^{-}-Cu^{2+}}$ of about $140^{\circ}$ is much larger than $90^{\circ}$ and rather close to $180^{\circ}$ for which strong antiferromagnetic superexchange interaction can be produced \cite{Kanamori}. 

Figure 3 shows the temperature dependences of total specific heat in $\mathrm{Cs_2Cu_3ZrF_{12}}$ and $\mathrm{Cs_2Cu_3HfF_{12}}$. Sharp peaks are observed at structural phase transition temperature $T_{\rm t}\simeq 210$ K and $175$ K for $\mathrm{Cs_2Cu_3ZrF_{12}}$ and $\mathrm{Cs_2Cu_3HfF_{12}}$, respectively. Peaks at $T_{\rm t}$ are not $\lambda$-like due to the first order transitions.
\begin{figure}[tbp]
	\begin{center}
		\includegraphics[width=15cm, clip]{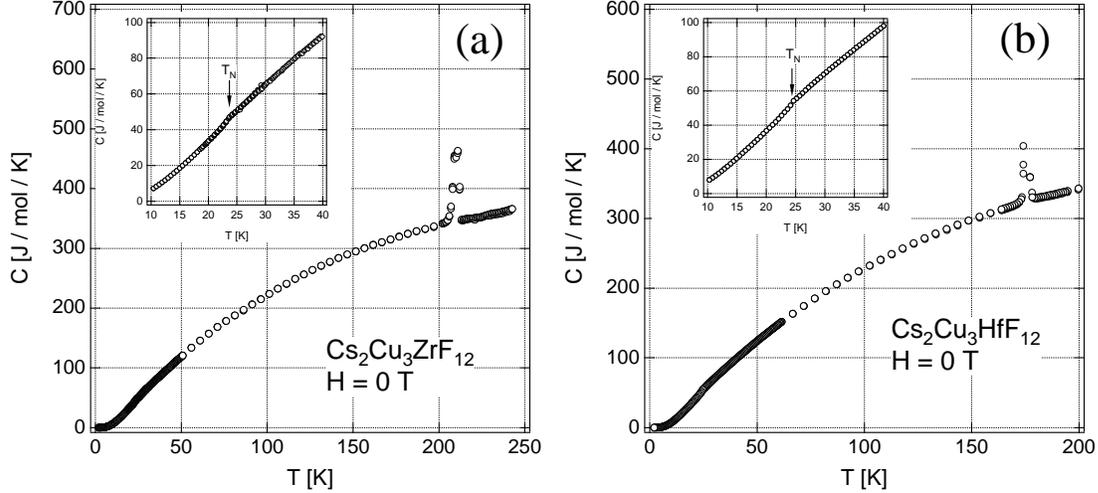}
	\end{center}
	\caption{Temperature dependences of specific heat in (a) $\mathrm{Cs_2Cu_3ZrF_{12}}$ and (b) $\mathrm{Cs_2Cu_3HfF_{12}}$ measured at zero magnetic field. Insets show the enlargements of low-temperature specific heat in both systems.}
	\label{heat}
\end{figure}

As shown in the insets of Fig. 3, the specific heat exhibits small bend anomaly at $T_{\rm N}=23.5$ K and 24.5 K for $\mathrm{Cs_2Cu_3ZrF_{12}}$ and $\mathrm{Cs_2Cu_3HfF_{12}}$, respectively. This anomaly is different from $\lambda$-like and cusplike anomalies which are characteristic of the phase transition of the second order. The small specific heat anomaly is indicative of small change in the entropy at $T_{\rm N}$. $T_{\rm N}$ increases with increasing applied field, e.g., for $\mathrm{Cs_2Cu_3HfF_{12}}$, $T_{\rm N}=28.8$ K at $H=9$ T. 
At present, the spin structure below $T_{\rm N}$ is unclear. The ratio of $|{\Theta}|$ to $T_{\rm N}$ is approximately 10 for both systems. The large $|{\Theta}|/T_{\rm N}$ value implies the good two-dimensionality and the strong frustration in the present systems.  
\begin{figure}[tbp]
	\begin{center}
		\includegraphics[width=12cm, clip]{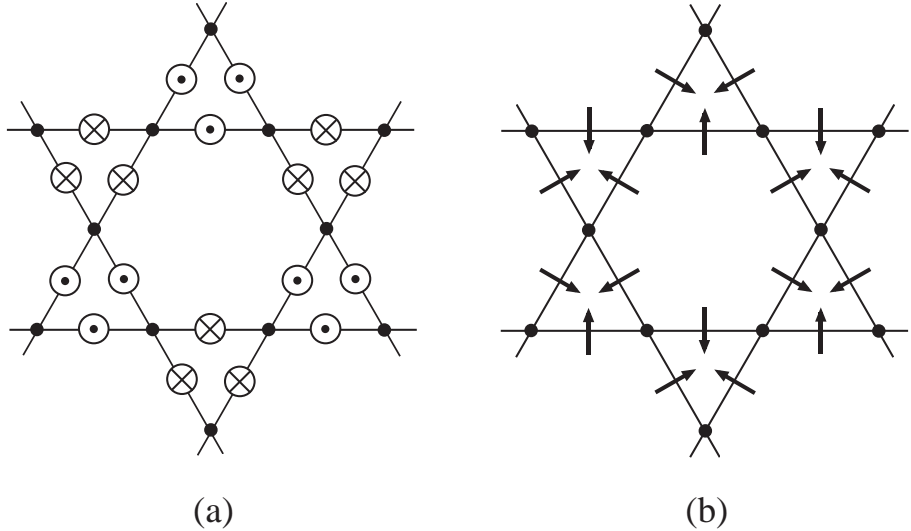}
	\end{center}
	\caption{Arrangement of the ${\bi D}$ vectors for $T > T_{\rm t}$ in $\mathrm{Cs_2Cu_3MF_{12}}$ systems, (a) the $c$-axis component and (b) the $c$-plane component.}
	\label{DM}
\end{figure}

Next, we discuss the origin of the weak ferromagnetic moment. There is no inversion center at the middle point of two neighboring magnetic ions in the Kagom\'{e} lattice. This situation differs from that in the triangular lattice. Thus, in general, the antisymmetric interaction of the Dzyaloshinsky-Moriya (DM) type, ${\cal H}_{\rm DM}=\sum_{\langle i,j\rangle} {\bi D}_{ij}\cdot [{\bi S}_i\times {\bi S}_j]$, is allowed. For the high symmetric crystal structure above $T_{\rm t}$ in which the regular Kagom\'{e} lattice is realized, there are mirror planes that are parallel to the $c$-axis, cross the middle points of neighboring Cu$^{2+}$ ions and are perpendicular to the lines connecting these ions. Therefore, the ${\bi D}$ vectors should be parallel to the mirror planes. Since there are two-fold screw axes along the $[1, 0, 0], [0, 1, 0]$ and $[1, 1, 0]$ directions, the ${\bi D}$ vectors become antiparallel along these directions. Thus, arrangement of the ${\bi D}$ vectors should be as shown in Fig. 4. In this case, if the Ne\'{e}l ordering occurs, the DM interaction acts to stabilize the so-called $q=0$ structure and also gives rise to the weak ferromagnetic moment due to the canting of ordered moments. If the ground state is spin liquid as RVB state, the DM interaction acts to mix the excited triplet state into the ground singlet state, which leads to the magnetic ground state with finite susceptibility. Although the details of the crystal structure below $T_{\rm t}$ is not clear at present, we infer that the weak ferromagnetic moment observed below $T_{\rm N}$ is attributed to the DM interaction, because the low-temperature crystal structure is expected to be closely related to the high-temperature crystal structure. 

\section{Conclusion}
We have presented the results of magnetization and specific heat measurements on the single crystals of $\mathrm{Cs_2Cu_3ZrF_{12}}$ and $\mathrm{Cs_2Cu_3HfF_{12}}$ which are described as quasi-2D $S=1/2$ Kagom\'{e} antiferromagnet. Both systems undergo structural phase transitions at $T_{\rm t}\simeq 210$ K and 175 K, respectively. Magnetic orderings accompanied with the weak ferromagnetic moment occur at $T_{\rm N}=23.5$ K and 24.5 K for $\mathrm{Cs_2Cu_3ZrF_{12}}$ and $\mathrm{Cs_2Cu_3HfF_{12}}$, respectively. At present, the spin structure in the ground state is not clear. The Dzyaloshinsky-Moriya interaction that is generally allowed in the Kagom\'{e} lattice should be responsible for the weak ferromagnetic moment. For further discussion on the magnetic ground state, we need detailed data of the crystal structure below $T_{\rm t}$.
The present systems are characterized as the $S=1/2$  Kagom\'{e} antiferromagnet with strong exchange interactions, $J/k_{\rm B}=360\sim 540$ K. Therefore, novel magnetic excitations predicted by recent theory\,\cite{Lecheminant,Waldtmann,Mambrini} may be examined by neutron inelastic scattering just above $T_{\rm t}$ where the regular Kagom\'{e} lattice is realized.

\ack
The authors would like to thank I. Yamada for useful suggestion on the sample preparation. This work was supported by a Grant-in-Aid for Scientific Research and the 21st Century COE Program at Tokyo Tech ``Nanometer-Scale Quantum Physics'', both from the Ministry of Education, Culture, Sports, Science and Technology of Japan.

\end{document}